\documentclass[a4paper,amsmath,amssymb,pre,twocolumn]{revtex4-1}
\usepackage{eepic}
\usepackage{graphicx}
\usepackage{subfigure}
\usepackage{latexsym}
\usepackage{bm}
\usepackage{amsmath}
\usepackage{amssymb}
\usepackage{overpic}
\usepackage{pstricks}
\usepackage{psfrag}

\def\k{\mathbf{k}}

\def\trho{\tilde{\rho}}
\def\ka{Kob-Andersen}
\def\bm{Biroli-M\'ezard}

\begin{document}

\author{Davide Cellai}
\altaffiliation[]{This work was carried out while DC was at the School of Chemistry and Chemical Biology, University College Dublin,
Belfield, Dublin 4, Ireland.}
\affiliation{Department of Mathematics and Statistics, University of Limerick, Limerick, Ireland}

\author{Andrzej Z.~Fima}
\author{Aonghus Lawlor}
\author{Kenneth A.~Dawson}

\affiliation{Centre For BioNano Interactions (CBNI),
School of Chemistry and Chemical Biology, University College Dublin,
Belfield, Dublin 4, Ireland}
%

\title{Crystallization in the vicinity of dynamical arrest}

\begin{abstract}
Dating from experiments more than 20 years ago, it has been realized that the crystallization of hard colloidal particles in the vicinity of dynamical arrest has several anomalies, that render the conventional nucleation and growth model inappropriate. Subsequently, key researchers have shown the influence of gravity. Here we show that a simple lattice model can capture most of the phenomena associated with such systems. 
In particular, the model reproduces not only characteristic signatures of glass-forming systems, but also the interplay between quasi arrested dynamics and crystal nucleation.
\end{abstract}

\date{\today}
\maketitle

\section{Introduction}
Dynamical arrest, that process in which many particles dramatically slow in a concerted manner, has been the focus of considerable attention for
some years \cite{mezard2000,dawson-cocis-2002,lawlor2005,ritort2003}. Packing-induced arrest, usually associated with repulsive
interactions between the particles, occurs because, as density increases, the amount of space available to a typical particle becomes so small that it
becomes effectively trapped by its neighbors \cite{
gotze1991,giovambattista2003}.
Even near arrest, though, rare particles in the system
may, as a result of fluctuations, have somewhat more empty space immediately available to them. If these rare spaces can be used for motion,
and the empty space generated by the motion passes throughout the system, then motion on long length scales can be generated.
Thus, there have been efforts
to describe dynamics using dynamically available volume, starting many years ago \cite{cohen1959}, and more recently a growing understanding of the
connection between available space and dynamics has begun to emerge \cite{degregorio2004,lawlor2005}.

However, there are problems related to, but comprising more than the issue of arrest itself, whose understanding would have broader scientific
impact across many disciplines. For example,  an understanding of the manner in which ordered structures grow in the vicinity of dynamical
arrest is one of the more pressing and important questions in the modern condensed matter science. In the early stages, nano-science greatly
focussed on the creation of increasingly small and more functional particles, and devoted proportionately less attention to rational approaches
to assembling structures from them. Now it is increasingly realized that useful devices will require us to fabricate ordered structures from
these particles, and in the longer term it will be necessary to approach these questions in a fundamental manner.
Research in arenas from photonic
crystals \cite{joannopoulos1995} to protein crystallization \cite{george1994,muschol1997,dawson-cocis-2002} has been hampered by the
fact that high quality ordered structures are hard to make, and may require specific, complex, and expensive methodologies. For the vast majority of
systems of practical interest, ordered structures compete with dynamical arrest leading to micro-crystallites in a matrix of glass, or partially
ordered materials with poor coherence, involving defects, dislocations, missing layers. Usually these phenomena degrade the functional properties
of the crystal, where they relate to technological issues \cite{starr2002}.

Hard particles should be somewhat simpler to understand. However, experimental studies of ordering kinetics of such particles, though began some years ago \cite{pusey1986, zhu1997}, are less understood than one might expect \cite{vanmegen1991}, given the
importance of the questions. The deficit of data is being rectified \cite{palberg1999,simeonova2004,schope2006,iacopini2009,tanaka2010}, but understanding of the existing results has been slow to develop, and many key issues are still not agreed within the community \cite{vanmegen1991,zhu1997,tanaka2010}.

The experimental results, mostly based on the model system PMMA and cis-decalin, are believed to be generic and are, in summary form, as follows.
As a function of increasing particle density, the hard-particle system exhibits first a single-phase fluid and then fluid-crystal co-existence.
Above the volume fraction $\phi\simeq0.545$ the pure crystal is the equilibrium phase. This crystal phase grows via homogeneous nucleation up to
$\phi_g\simeq 0.58$, but beyond this there appears to be a sharp cross-over to (very slow) heterogeneous nucleation, mainly from the edge
of the vessel, and free surface \cite{pusey1986}.
Contrary to normal expectation, the size of the crystallites formed decreases as one goes deeper into the crystal region (higher volume fraction) and one approaches arrest.
This sharp phenomenon has been understood by many authors to be itself dynamical arrest, and theoretical treatments of it have been generally considered successful \cite{pusey1987,gotze1991} though these observations are complicated by new findings.
Indeed, gravity also seems to play a role in hindering crystallization \cite{zhu1997,simeonova2004}.
Moreover, recent molecular dynamics simulations show that even in the absence of diffusion crystallization is possible in bulk homogeneous monodisperse hard-spheres \cite{zaccarelli2009}.

In our approach we would like to probe directly the emergence of dynamically arrested substances in a system that is also able to crystallize.
Our aim, therefore, is to deal with both arrest and crystallization in a fully consistent manner and establish the connection between the two.
We would also like to see how ordering phenomena are controlled by the numbers and heights of barriers to particle movement (on all length scales) that originate from the caging and to understand how (rare) empty space is managed in the process of forming ordered structures.

\section{The model}

\begin{figure}[htb]
\begin{minipage}[b]{0.49\columnwidth}
    \centering
    \subfigure[]{%
        \label{fig:subfig:caging}
        \includegraphics[width=\columnwidth,angle=0]
        {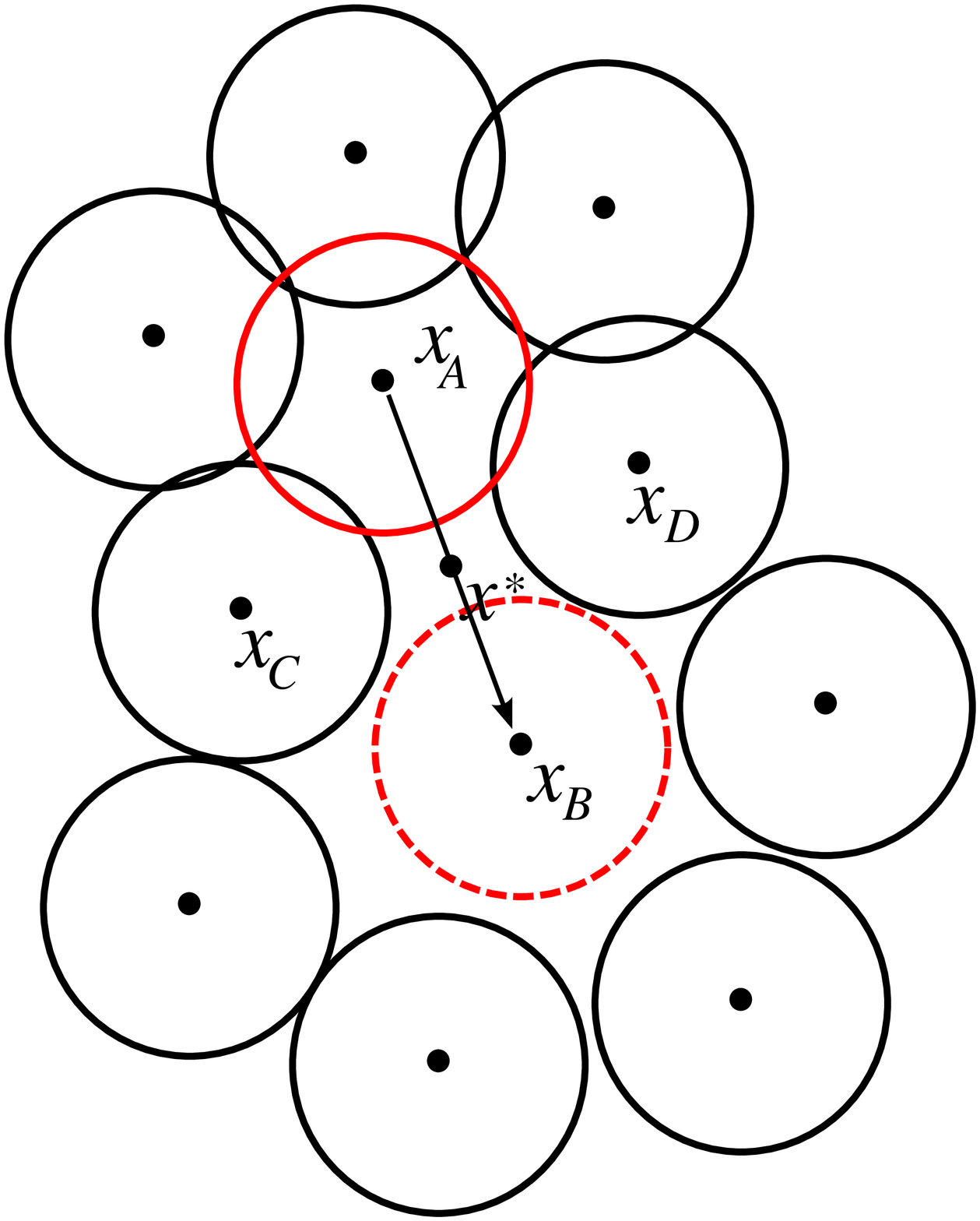}%
    }%
\end{minipage}%
\begin{minipage}[b]{0.49\columnwidth}
    \centering
    \subfigure[]{%
        \label{fig:subfig:barrier}
        \includegraphics[width=\columnwidth,angle=0]
        {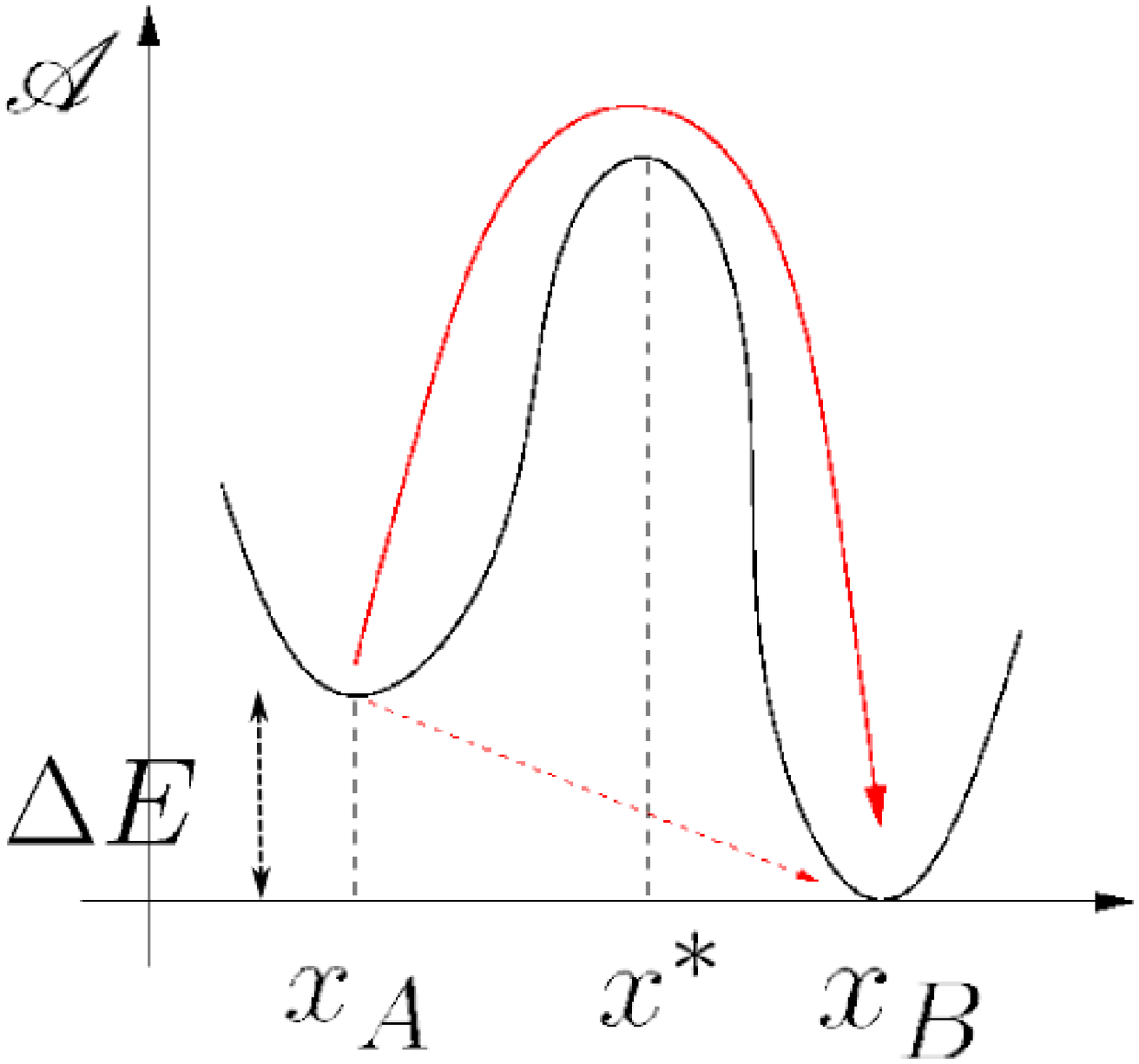}%
    }%
\end{minipage}
\caption{
(a) Schematic representation of the caging phenomenon. 
		To go from a crowded ($x_A$) to a less crowded position ($x_B$), the particle has to overcome the barrier (b) due to particles in the immediate vicinity (such as $C$ and $D$).
    Cage escape rates, that generally depend on the type of interaction and particle density, are represented by kinetic rates in Kob-Andersen models \cite{kob1993}.
    $\Delta E$ is the difference in local free energy between cages.
}
\label{fig:ka_meaning}
\end{figure}

Our model has been first introduced in Ref.~\cite{cellai2010} to investigate the dynamics of glass-forming systems in the presence of crystallization.
The model represents two phenomena typical of extended particles in a crowded environment, and we shall explain it with the help of Fig.~\ref{fig:ka_meaning}.
The first aspect is that  particles will tend to move from more to less crowded portions of space (see Fig.~\ref{fig:subfig:caging}) in order to mimimize their local free  energy.
The second aspect is that they may have to wait for sufficient kinetic energy to cross a local barrier, or wait for some (microscopic) time period for a neighboring particle to move aside in order to access a nearby space. Both of these effects can be represented by a barrier in the local free energy (Fig.~\ref{fig:subfig:barrier}).
The aim is then to understand how these microscopic energy scales lead to long length and time scale behaviors.

First, we define a Hamiltonian which makes the more crowded environments energetically relatively unfavourable:
\begin{equation}
 H = V_R \sum_{j=0}^{V}  \,(n_j - c_R)\, \theta(n_j - c_R).
 \label{eq:hamiltonian}
\end{equation}
Here $c_R$ is the maximum number of nearest neighbors that may
surround a particle without it incurring a cost, $n_j$ the number of nearest neighbors of the particle at the $j$-th site and $V_R$ is the
strength of the repulsive interaction, $\theta(x)$ is the Heaviside function and $V$ the total number of sites (volume).
This Hamiltonian can be seen as an extension of the {\bm} model, where point particles on a lattice are forced to keep some exclusion volume around them \cite{biroli2002}.
Such a type of interaction mimics, on a lattice, the behavior of hard spheres and therefore originates a crystal phase.
In this paper we consider a soft repulsion to be able to mimic slightly soft spheres on a lattice \cite{mccullagh2005}.

The second phenomenon we represent is the so called \emph{local caging} effect, in which the locations of the neighbors of a particle lead to a local free energy barrier to access a neighboring empty space.
It has been shown by experiments \cite{weeks2000}, and also reproduced by continuum simulations \cite{auer2003}, that dense colloids are characterized by this phenomenon.
A particle spends most of the time rattling inside the cage formed by its neighbors and occasionally, if a fluctuation of the neighbors opens up the cage, it makes a longer movement and starts rattling again in another cage.
As schematically shown in Fig.~\ref{fig:subfig:caging}, the particle  is caged by several neighbors in such a way that the illustrated movement is only possible if the original cage of particles fluctuates and opens sufficiently to provide an exit path.
The relative unlikelihood of this kind of fluctuations originates in the fact that there are relatively few such configurations.
The local free energy will therefore reflect this in having a barrier between these two adjacent local minima (Fig.~\ref{fig:subfig:barrier}).
The barrier energy scale is represented by a rate constant for single particle motion \cite{kob1993,jackle1994,ritort2003} which can be implemented in a Monte Carlo scheme.
We therefore define the following kinetic rule.
A particle can move from a site $i$ to one of its empty nearest neighboring sites $j$ only if:
\begin{itemize}
 \item[(a)] the sum of its nearest and next-nearest neighbors is not larger than a fixed parameter $c_K$;
 \item[(b)] the movement is reversible, \textit{i.e.} the particle in $j$ can go back to site $i$ without breaking the rule (a).
\end{itemize}
This rule takes into consideration the intuition that the number of particles involved in the local cage is larger than the one which controls the amount of space available for intra-cage motion.
Thus, we assume that the arrangement of particles involved in a local cage extends up to the next-nearest neighbors.
Moreover, because of the discrete nature of the model, a wider range of values of $c_K$ allows a more satisfactory fine tuning of the kinetic constraint.

This kinetic rule has been  inspired by the Kob-Andersen model \cite{kob1993}, which reproduces many  signatures of systems close to dynamical arrest, including blocked non-ergodic states and dynamical heterogeneities \cite{lawlor2005,pan2005}.
Moreover, it has been understood that empty spaces of a highly dense lattice model can be categorized according to their role in the dynamics.
Holes are defined as empty sites which a particle can move into and constitute the dynamically available volume of the system \cite{lawlor2005}.
In a dense configuration, most holes will be only involved in local movements.
However, for some of the holes there may be at least one sequence of movements which allow the hole to move almost all the particles in the system.
Such holes are named \emph{connected holes} and play a fundamental role in determining the diffusivity of the system \cite{degregorio2005,lawlor2005}.
An important characteristic of this class of kinetically constrained models is the presence, in addition to the local cages, of large cages, \textit{i.e.} extended closed arragements of particles which prohibit every particle inside the cage to move outside.
Only from rearrangements of the particles on the outside those cages can be broken \cite{degregorio2004}.
The presence of extended cages in our model has certainly an effect on the process of crystallization that we want to investigate.
 
The two previous features of the model can be implemented in a Monte Carlo scheme by defining the total transition rate probability $P_{i\to j}$  of a particle going from site $i$ to site $j$ as the product of a kinetic term and an energy term:
\begin{equation}
P_{i\to j} = K_{i\to j} F_{i\to j}.
\end{equation}
The energy term is the usual Metropolis rule
\begin{equation}
F_{i\to j} = \min\{e^{-\Delta E},1\},
\end{equation}
where $\Delta E$ is the difference in energy of the two states according to the Hamiltonian (\ref{eq:hamiltonian}).
The kinetic term, which implements the constraint we have discussed above, is
\begin{equation}
K_{i\to j} = \theta(c_K-n_i)\theta(c_K-n_j),
\end{equation}
where $n_i$ and $n_j$ are the sums of nearest and next-nearest
neighbors of the site $i$ and $j$, respectively, and $\theta(x)$ is the Heaviside function, with the convention $\theta(0)\equiv 0$.
The quantity $T\equiv\beta^{-1}$ thereby represents the effective height of the local barrier.

\begin{figure}
\begin{center}
    \includegraphics[height=\columnwidth,angle=270]{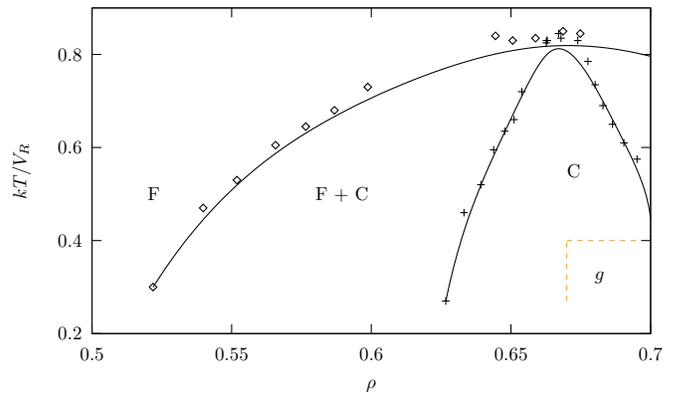}
\end{center}
\caption{
		Section of the equilibrium phase diagram of the model for $c_R=3$ on a cubic lattice \cite{mccullagh2005}, $F$=fluid, $C$=crystal.
    In the crystal phase for $T\lesssim 0.4$ and $\rho>2/3$, signatures of a glass state ($g$) are observed for $c_K=10$ \cite{cellai2010}. }
\label{fig:phasediag}
\end{figure}

\section{Properties of the model}

In this paper, we want to focus on aspects related to crystallization and nucleation close to arrest.
Evidently such a model has an extensive parameter space, which is indeed likely to represent a wide range of physically realizable situations.
However, here we wish to explore a scenario that is close to the widely studied hard sphere case.
Choosing $c_R=3$, $V_R=1$, it is possible to calculate an equilibrium phase diagram which presents the required characteristics.
A description of the equilibrium phase diagram \cite{mccullagh2005} for such parameter values, as well as a detailed study of the dynamical properties \cite{cellai2010} of this model have been given elsewhere.
For the reader's convenience, we briefly summarize those results.
Fig.~\ref{fig:phasediag} reproduces an interesting section of the phase diagram.
In that section, when the temperature is low enough, the phase diagram displays the sequence of phases (for increasing density: liquid, liquid-crystal coexistence, crystal), driven by repulsive short-ranged interaction, as in colloids.
For $c_R=3$, the crystal of our model consists of double diagonal layers alternated with single empty diagonal layers and therefore it has a periodicity of $\sqrt{3}$ lattice steps.

The origin of the fluid-crystal phase transition in this model can be explained by an entropic argument.
The short-ranged repulsion represented by the Hamiltonian (\ref{eq:hamiltonian}) has the effect of imposing some excluded volume in the system.
Therefore, in the supercooled region the ordered crystalline structure is entropically favoured respect to the disordered fluid phase, due to the more efficient use of the free spaces, as in hard-sphere systems \cite{frenkel1999}.
The onset of crystallization in the supercooled region is associated with the appearance of local assemblies of particles with the correct crystal structure.
Such assemblies are thermodynamically stable and we will refer to them as crystal nuclei.

Inside the crystal phase, when we set $c_K=10$, an apparent dynamical arrest for density $\rho\gtrsim 0.66$ and $T\lesssim 0.4$ is observed (schematically indicated in Fig.~\ref{fig:phasediag}).
In this region, several signatures can be found, such as extremely slow energy relaxation and non-exponential slowing phenomena, with good fitting of the Kohlrausch-Williams-Watts law.
The calculation of the Kauzmann temperature yields $T_K=0.42$ \cite{cellai2010}.

\begin{figure}
\begin{center}
    \includegraphics[width=0.80\columnwidth,angle=0]
    {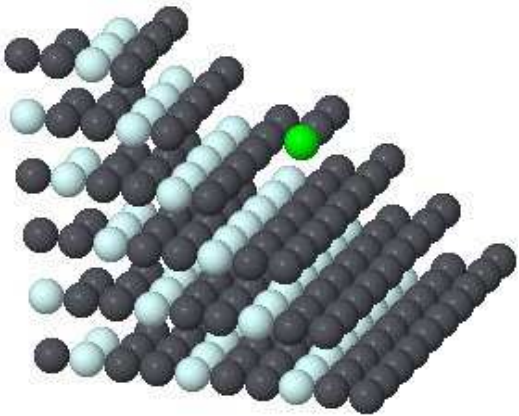}
\end{center}
\caption{Scheme of sequential growth of the crystal for $c_R=3$.
	In order to insert the particle marked in green, the kinetic
	parameter $c_K$ has to be larger or equal to $6$.}
\label{fig:growth}
\end{figure}

\begin{figure}
    \includegraphics[height=0.90\columnwidth,angle=270]{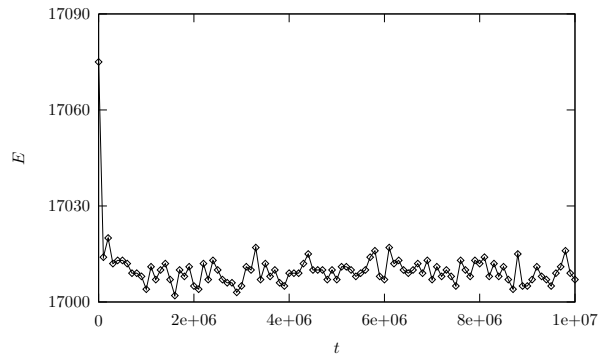}
\caption{Energy evolution in number of lattice sweeps of a simulation at $c_K=6$, $T=0.4$ and $\rho=0.64$.
This energy behavior seems to indicate a steady state. Actually, the configurations are disordered, but the equilibrium state is a crystal here.
Therefore, the system is in a long-lived non-equilibrium state.}
\label{fig:simulation_ck6}
\end{figure}

This glassy behavior is determined by the presence of the kinetic rule, as in its absence the system quickly crystallizes \cite{mccullagh2005}.
The role of the parameter $c_K$, in fact, is mainly to control the barriers and consequently both dynamics and kinetics.
Firstly, the value of $c_K$ is greatly involved in the detailed means by which crystals are formed.
The sequential nature of the growing of layers or the ease with which one can fill a defect are features mainly governed by $c_K$.
Let us consider the crystal for the case $c_R=3$ and $V_R=1$ at $T=0$ (i.e.~in the hard sphere case).
It is interesting to note that a small value of $c_K$ can determine quite strictly the mechanism by which the crystal can form.
The minimum number of neighbors (i.e. the sum of nearest- and next-nearest-neighbors) which is necessary to fill a point defect into an isolated single diagonal layer of particles is only $3$, but in order to fill a defect in a \emph{double} diagonal layer, the number is $9$.
A realistic scheme for growing a crystal at high densities is
the case of a sequential growth on the border of an
incomplete layer which is adjacent to a full layer (see Fig.~\ref{fig:growth}).
In this case, the lowest value of $c_K$ for growing the crystal is $6$.
However, $c_K=6$ still appears to be a too low value to show an appreciable degree of diffusivity.
As an example, Fig.~\ref{fig:simulation_ck6} shows the energy evolution of a long simulation for $\rho=0.64$  at $T=0.4$.
According to Fig.~\ref{fig:phasediag}, this point is close to the left border of the crystal phase, but no crystalline state is observed.
On the contrary, it is worth to remark again that we want to reproduce, as well as it is possible in a lattice model, the experiments which display classical crystallization and homogeneous nucleation as well as slow dynamics and arrest \cite{pusey1986}.
In that respect, the behavior at $c_K=10$ is more interesting.
Here two distinct dynamical regimes (crystallization and arrest) are easily identifiable.
This observation suggests that in the process of nucleation the sequential growth is probably not enough to guarantee the crystallization.
The possibility of filling point defects and, more generally, of growing layers without a prescribed order plays a crucial role.
Therefore, in the following results we are going to  study extensively the model defined by $c_K=10$.

It is of some interest to understand the onset of this apparent dynamical arrest transition as temperature is lowered, and density increased,
and the simplicity of the model makes it possible to dissect the mechanisms.
For such models, changes in the nature of the dynamical processes (essentially the breakdown of the Stokes-Einstein relation) have been understood in terms of a dramatic reduction of the number of easy pathways in the phase space \cite{lawlor2005}.
In turn this has been related to the manner in which excess space is organized, connected empty space allowing for long range dynamical process, and diffusion.
Thus, if one creates an isoenergetic phase space ({\ka} model) from the present model, by removing the energy term, the resulting model of the homogeneous fluid exhibits a violation of the Stokes-Einstein relation.
When the energy terms are restored, the number of free motions in  the system is reduced yet further and, just beyond the Stokes-Einstein violation, the system
appears to arrest.
Nevertheless, one should not assume that this is necessarily a true glass transition, for the divergence of characteristic time is obtained by extrapolation, and there remains slow, but visible motions well beyond the apparent arrest.
However, we can say with certainty that in this region the system becomes sub-diffusive, and the slowing is so dramatic that it is hardly possible to discern the difference from a real glass \cite{cellai2010}.


We observe that, for sufficiently high barrier heights, this phase diagram is similar to that shown in many experimental studies of the dense hard sphere system \cite{pusey1986,vanmegen1993}.
Experiments also show that the kinetics of crystallization becomes different in approaching the glass transition.
In the next section we seek to simulate those experiments using our model, as a possibility to check the quality of the mathematical representation and to give new insight into the phenomena involved.


\section{Kinetics of crystallization}
We are now in position to investigate in the model the phenomena associated with crystallization.
Here the aim is to understand to what degree the model reproduces nucleation and growth, and if it is possible to recognise dynamical arrest as in the experiments of colloidal particles.
In our analysis we will often refer to the two important characteristics of the model, namely the fluid-crystal phase transition generated by the short-ranged repulsion and the properties of the kinetic constraint, including the presence of extended cages in the system.
Finally, we will try to understand if the model can even predict laws that could be checked experimentally.

\subsection{Configurations}
\begin{figure*}[htb]
\begin{center}
	\begin{minipage}[b]{0.32\textwidth}
					\subfigure[$\rho=0.580$]{%
									\includegraphics[width=0.99\columnwidth,angle=0]
									{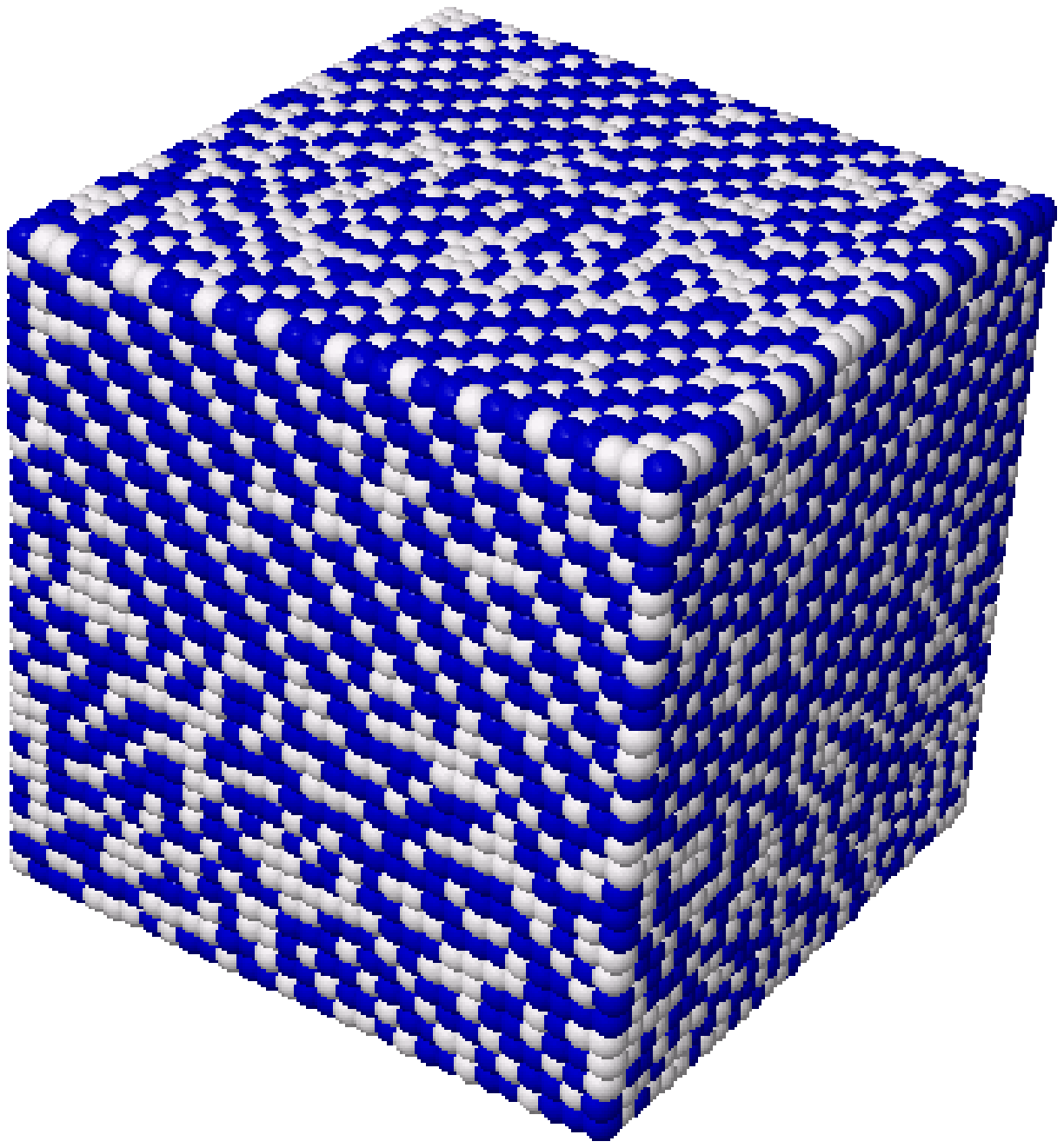}%
					}
	\end{minipage}%
	\begin{minipage}[b]{0.32\textwidth}
					\subfigure[$\rho=0.640$]{%
									\includegraphics[width=0.99\columnwidth,angle=0]
									{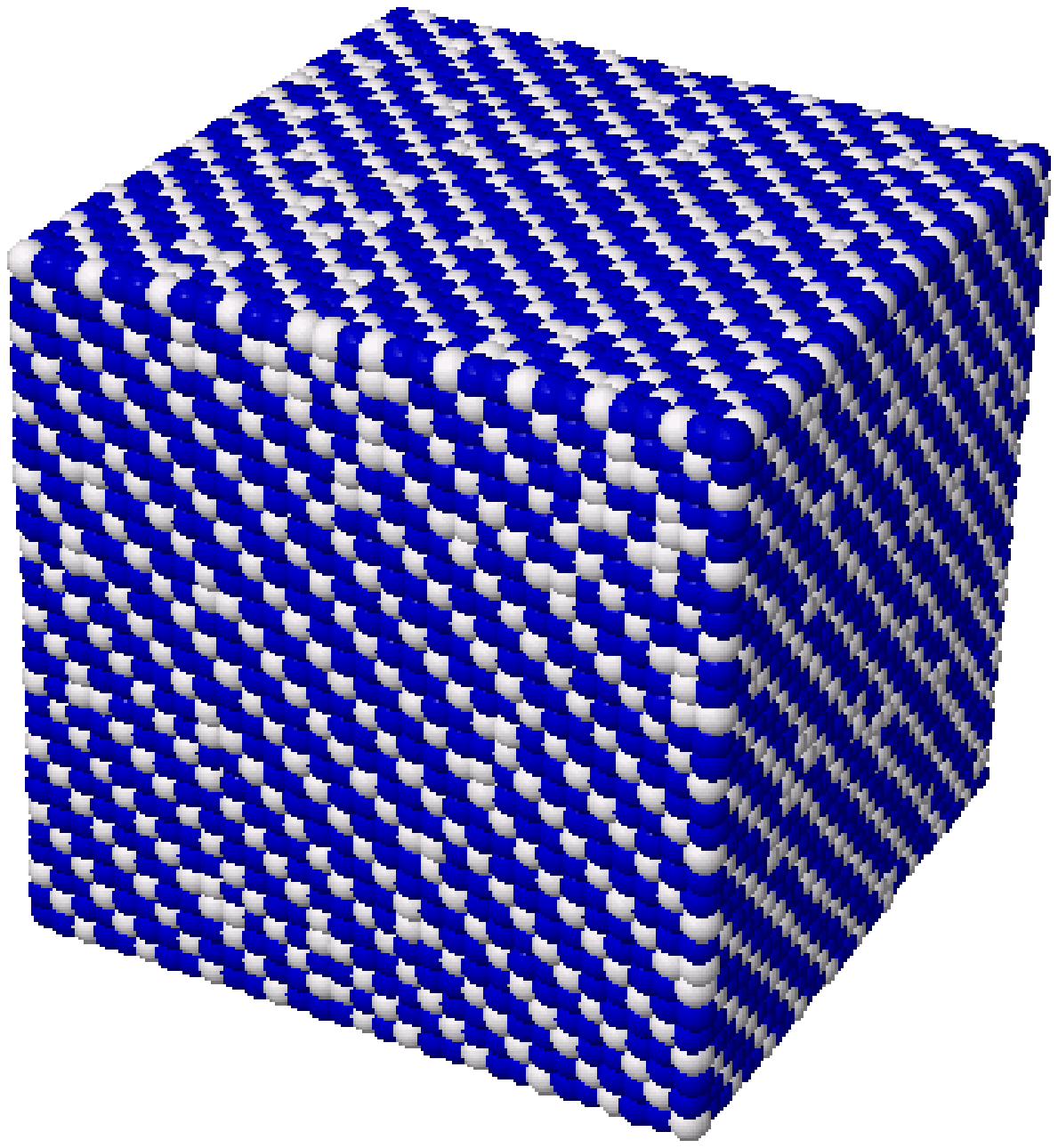}%
					}
	\end{minipage}%
	\begin{minipage}[b]{0.32\textwidth}
					\subfigure[$\rho=0.645$]{%
									\includegraphics[width=0.99\columnwidth,angle=0]
									{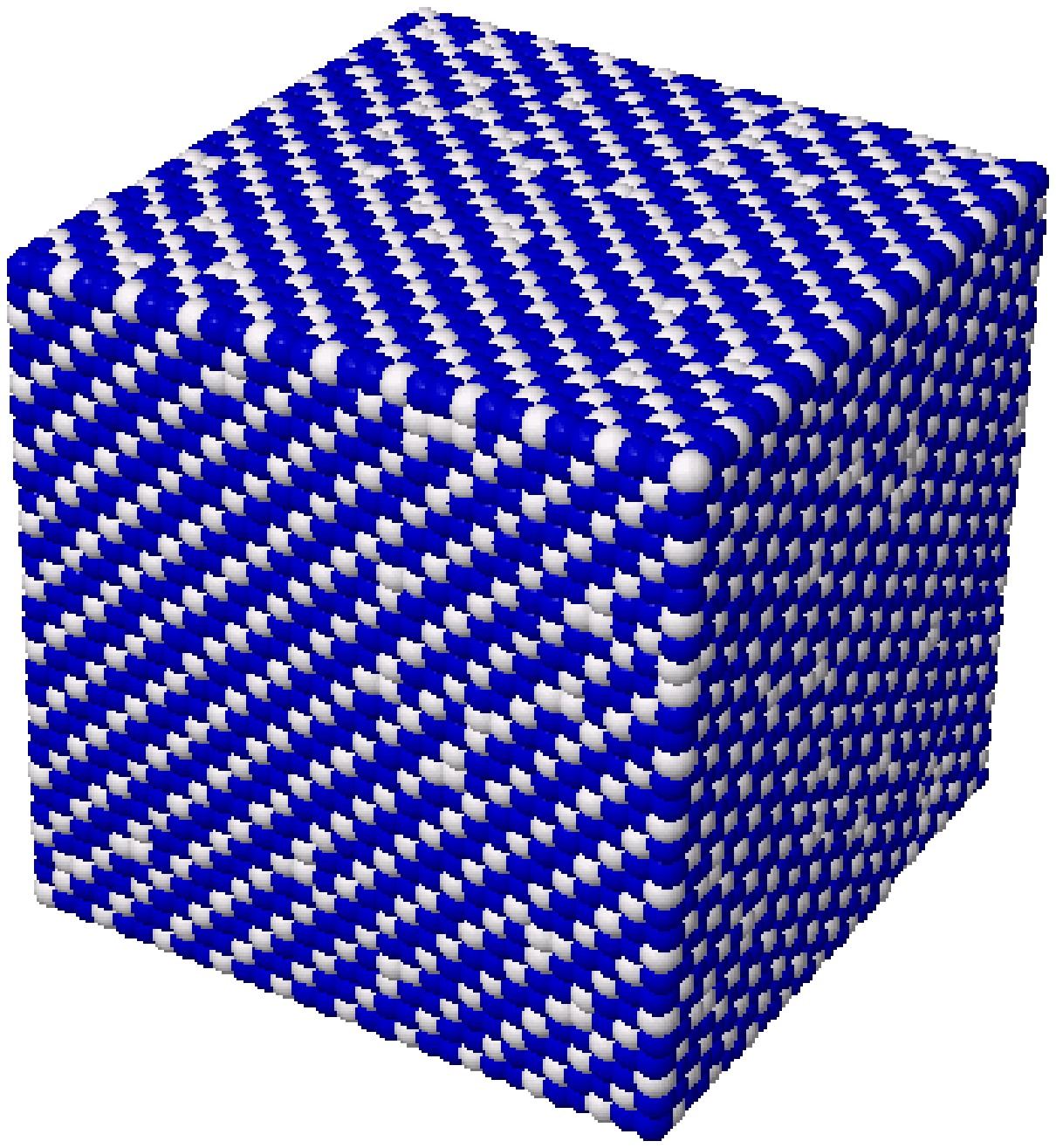}%
					}
	\end{minipage}%
	
	\begin{minipage}[b]{0.32\textwidth}
					\subfigure[$\rho=0.650$]{%
									\includegraphics[width=0.99\columnwidth,angle=0]
									{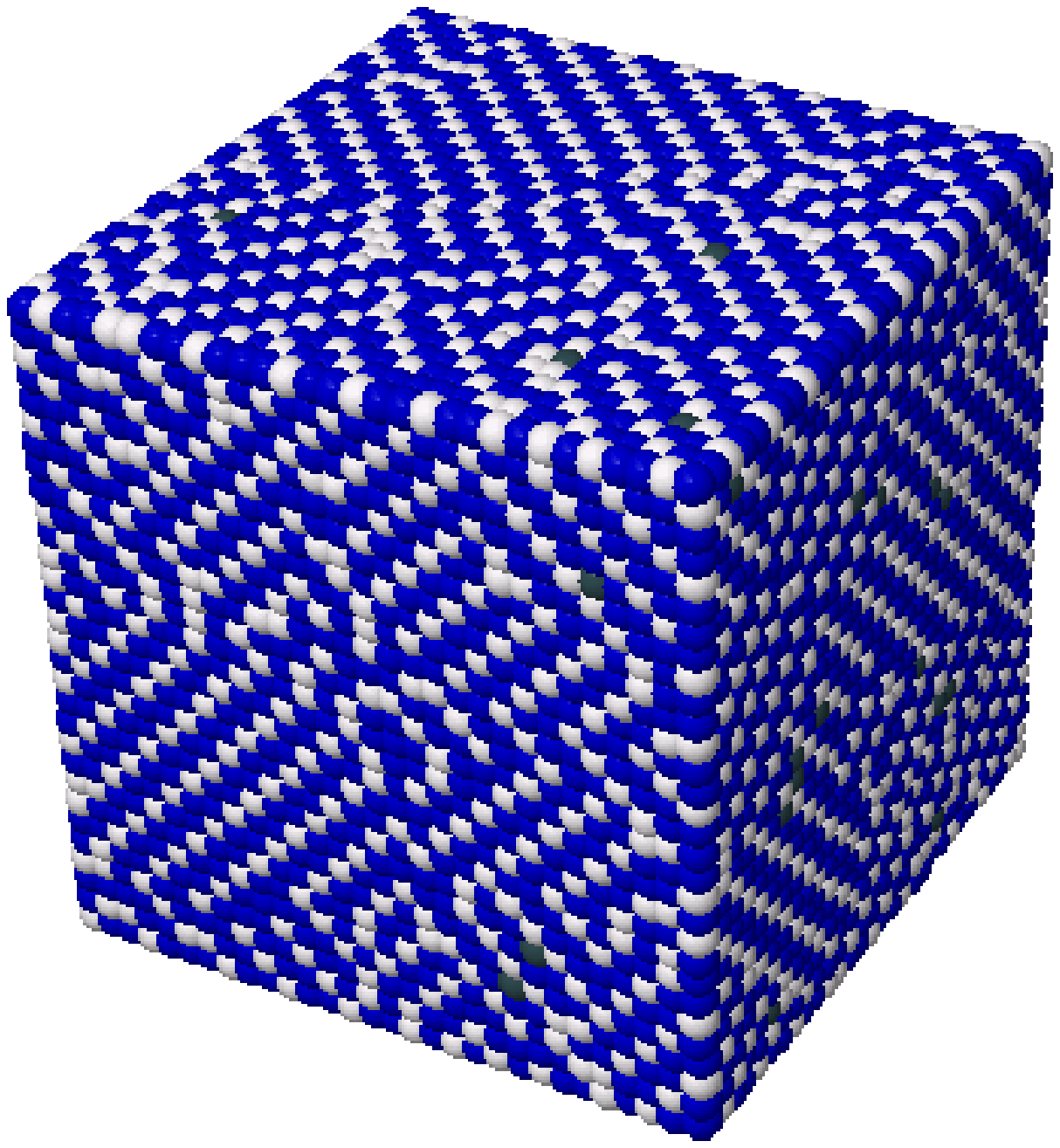}%
					}
	\end{minipage}%
	\begin{minipage}[b]{0.32\textwidth}
					\subfigure[$\rho=0.655$]{%
									\includegraphics[width=0.99\columnwidth,angle=0]
									{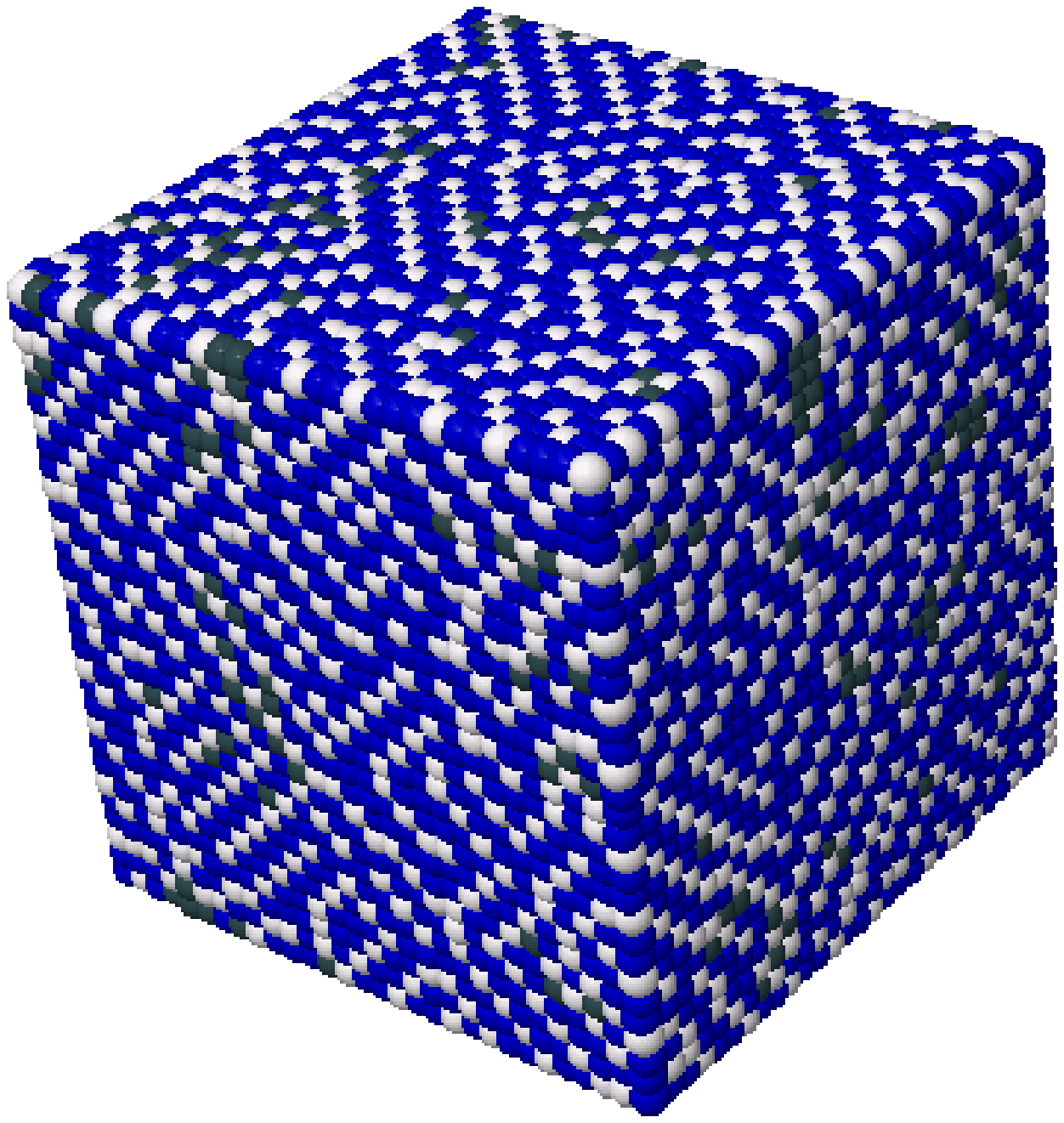}%
					}
	\end{minipage}%
	\begin{minipage}[b]{0.32\textwidth}
					\subfigure[$\rho=0.665$]{%
									\includegraphics[width=0.99\columnwidth,angle=0]
									{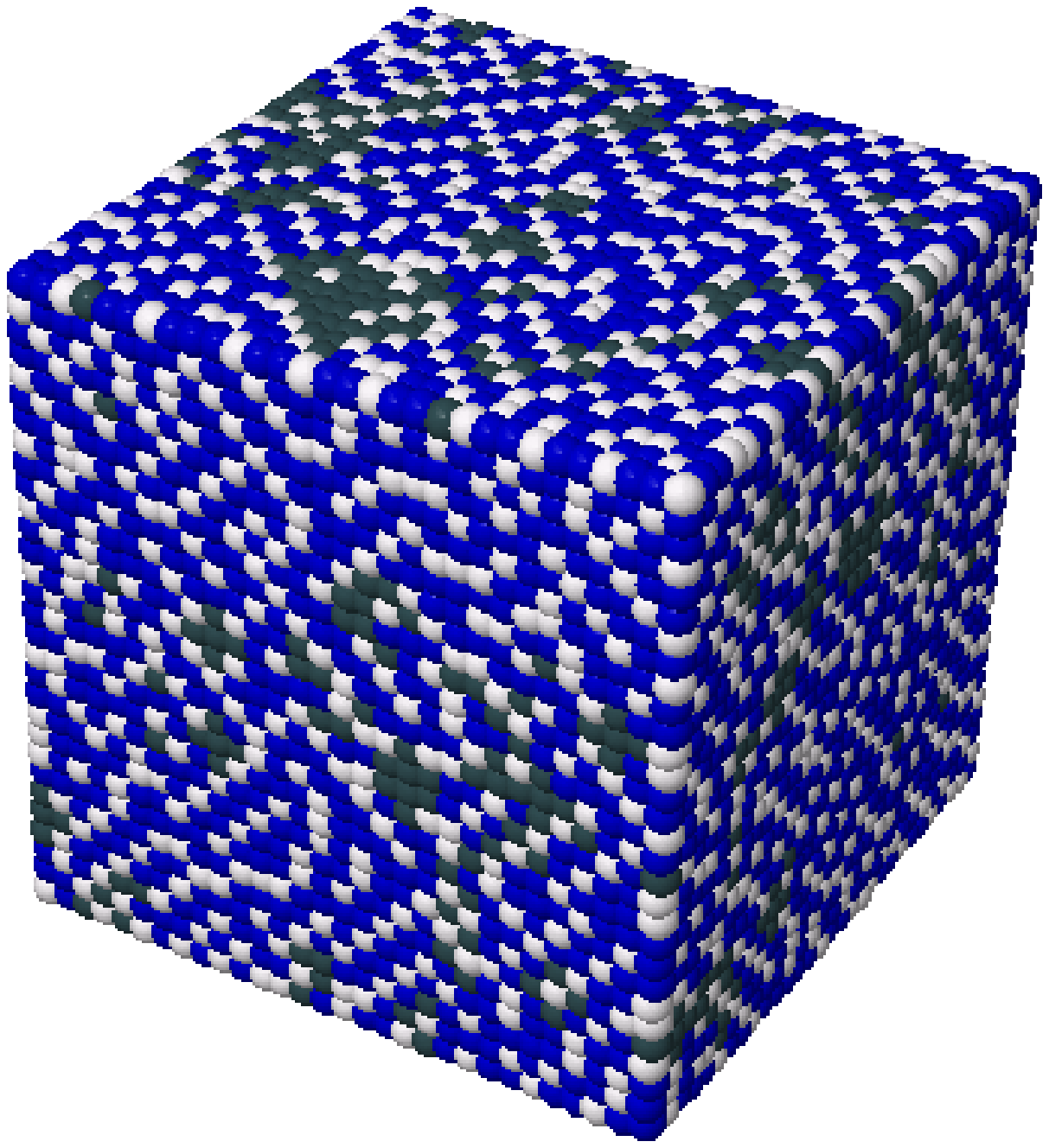}%
					}
	\end{minipage}
\end{center}
\caption{
Simulation of the experiment by \cite{pusey1986}.
	Configurations in the fluid state at a given density $\rho$ are instantaneously quenched at $T=0.4$, where the equilibrium state is crystalline.
	Snapshots of the system after a relaxation process of $3\cdot 10^5$ Monte Carlo steps are presented.
	Blue spheres represent particles that have moved in the last $10^4$ time steps.
	Black spheres are particles which have not recently moved.
}
\label{fig:states}
\end{figure*}

To begin with, we phenomenologically illustrate the kinetics of relaxation by showing configurations of different densities as a function of time.
After a period of equilibration, we see, depending on the system density, fluid-crystal, crystal and ``arrested'' regimes as the density of the system is increased.
This is shown in the sequence of figures (Fig.~\ref{fig:states}) where  snapshots have been taken after a ``long'' period of time: meaning that all of the fast processes have ceased, and only slow aging remains.
This representation may be directly related to the samples shown in the original papers \cite{pusey1986,vanmegen1993}.
For the lowest density ($\rho=0.58$), we see, as expected from the equilibrium phase diagram, phase separation between fluid and crystal.
Similarly, the next few densities quite rapidly equilibrate to the pure crystal state.
However, with increasing density the system forms smaller crystallites that jam each others subsequent progress.
Only after much larger times these crystallites coarsen to form a single crystal.
Near the apparent arrest ($\rho=0.655$) the size of the crystallites remains quite small.
However, it must be pointed out that the samples at $\rho=0.655$ and  $\rho=0.665$ will eventually crystallize after a longer time.
As it was observed in the previous section, there are several indications that the hypothetical arrest should occur at $\rho=2/3$, which is the density of the perfect crystal.
Beyond $\rho\gtrsim 0.67$ the system is slower and slower and the resulting states are more and more disordered at a fixed time.

\begin{figure*}[htb]
\begin{center}
	\begin{minipage}[b]{0.15\textwidth}
		\subfigure[$t=0$]{%
			\includegraphics[width=0.99\columnwidth,angle=0]
			{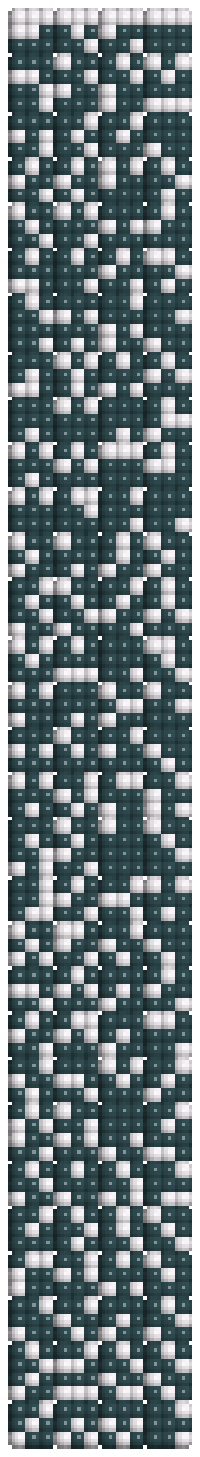}%
		}
	\end{minipage}
	\begin{minipage}[b]{0.15\textwidth}
		\subfigure[$t=3\cdot 10^5$]{%
						\includegraphics[width=0.99\columnwidth,angle=0]
						{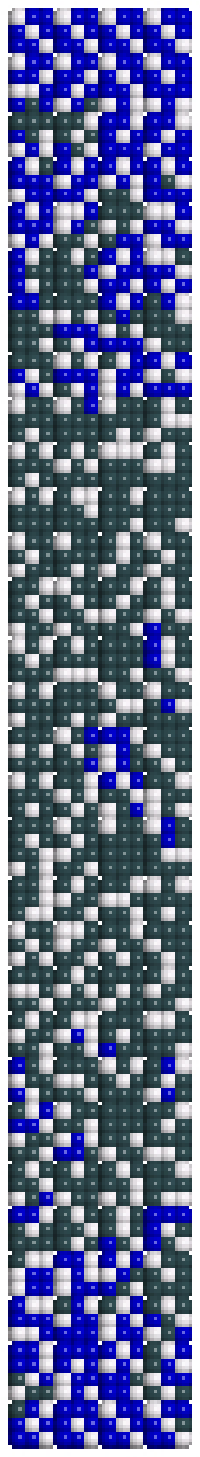}%
		}
	\end{minipage}
	\begin{minipage}[b]{0.15\textwidth}
		\subfigure[$t=6\cdot 10^5$]{%
						\includegraphics[width=0.99\columnwidth,angle=0]
						{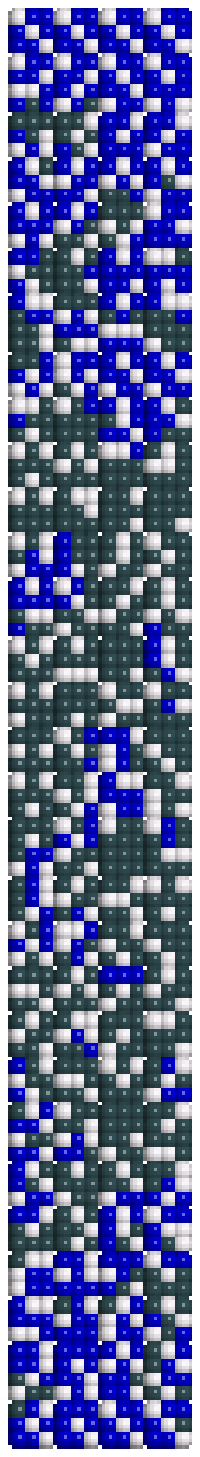}%
		}
	\end{minipage}
	\begin{minipage}[b]{0.15\textwidth}
		\subfigure[$t=10^6$]{%
						\includegraphics[width=0.99\columnwidth,angle=0]
						{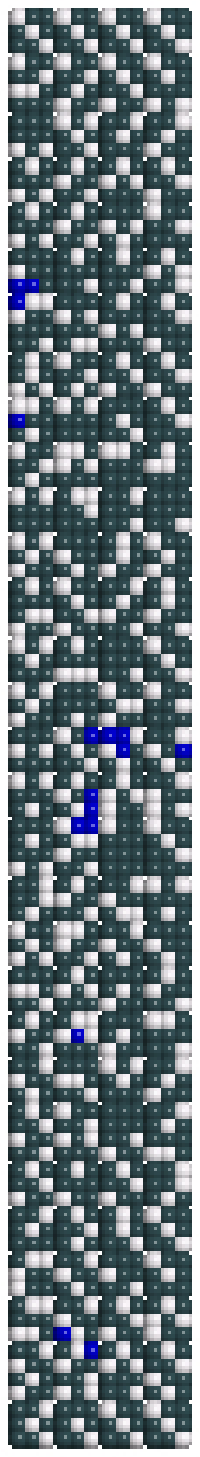}%
		}
	\end{minipage}
	\end{center}
	\caption{
	Simulation of heterogeneus nucleation. 
	A layer of particle is removed from the top of an ``arrested'' configuration at density $\rho=0.69$ (the resulting density after the removal  is $\rho=0.683$).
	Snapshots of size $12\times12\times96$ are shown during the relaxation process.
	Due to the extra free space, the system partially crystallizes.
	After $t\approx 6\cdot 10^5$, the number of movable (blue) particles  appear to decrease quite sharply.
	Blue and black spheres have the same meaning as in Fig.~\ref{fig:states}.
	We always use periodic boundary conditions.
	}
\label{fig:removing_layer}
\end{figure*}

In Figure \ref{fig:removing_layer}, another type of analysis is presented.
We consider a glass state at high density ($\rho=0.69$) and remove a layer.
In order to reduce as much as possible the finite size effect, we consider an elongated sample.
In this way, the removal of a layer on the small side perturbates the system less, because it reduces the density decrease, so that the characteristics of the sample are not profoundly changed.
Due to the excess of available empty space, the system partially crystallizes in the proximity of the emptied layer.
However, the formed crystallites  appear to stop growing at about $t\sim 6\cdot 10^5$.
In fact, the number of moving particles decreases sensibly and in the centre of the sample we observe a block of matter which appears quite resilient to movement.


\subsection{Analysis of the static structure factor}
In order to discuss more precisely these qualitative remarks, we study the time evolution of  the static structure factor $S(k)$, which can be regarded as an indication of the crystallinity of the system.
We calculate the static structure factor of the configurations generated by our simulations using the formula
\begin{equation}
	S(\k) = \frac{1}{N}\langle\trho_{\k}\trho_{-\k}\rangle,
\end{equation}
where the Fourier components $\trho_{\k}$ of the density are calculated by the Fast Fourier Transform algorithm.

A few comments should be made about the meaning of the structure factor in a lattice model.
On a lattice, all the coordinates and the distances between pairs of points are discrete.
For example, this implies that the number of pairs at short distances $r$ is necessarily smaller than at larger distances.
This causes a poor sampling average in calculating the spherically-symmetric integrated structure factor $S(k)$ at large $k$.
Moreover, the behavior of $S(k)$ in a crystal is quite different in a lattice model from a continuum model or experimental plots.
Here the static structure factor of a perfect crystal consists of a single spike at the value of $k$ corresponding to the periodicity of the crystal, and the peak has no width.
On the contrary, in a real system molecules oscillate around their equilibrium positions and therefore slight changes in the measured periodicity occur.
On the other hand, the structure factor of an imperfect crystal on a lattice presents some secondary peaks close to the main crystal peak, but this does not mean that they represent an additional periodicity in the sample.
They can rather be considered as discrete effects of the width of the main peak, due to defects, dislocations or broken layers.
In a real system, there is a much larger number of particles and it is therefore highly probable that  all  possible slight modifications of the correct periodicity are present.
Therefore, we conclude that in a lattice model the height of the main peak of the structure factor quantifies more meaningfully the degree of order in the system than an integration over a number of close peaks.

This is also the reason why we do not implement the definition of crystallinity recently used to analyze experimental data \cite{schope2006,iacopini2009}, where the crystallinity is defined as the integral of the static structure factor over the peak associated with the relevant periodicity.
Moreover, it must be stressed that here we are not interested in the structure factor itself, but in the divergence of the crystallinity in approaching dynamical arrest.
As in the vicinity of arrest the growth of crystal nuclei involves a very long sequence of correlated movements (due to the increasing density), the region of space involved in the crystal growth becomes larger and larger, so that the divergence of the characteristic times of the process cannot depend very much on the local details of the formed ordered zones in the sample.
Proper crystallinity  measures the amount of order in the sample, whereas the peak of the structure factor simply represents an estimate of the size of the perfect crystal.
We assume that these two quantities diverge with the same law in approaching dynamical arrest.
Finally, we point out that the choice of using the height of the peak of the structure factor has been made in other experiments described in the literature \cite{schatzel1993}.

\begin{figure}[hbtp]
\begin{center}
	\includegraphics[height=0.99\columnwidth,angle=270]{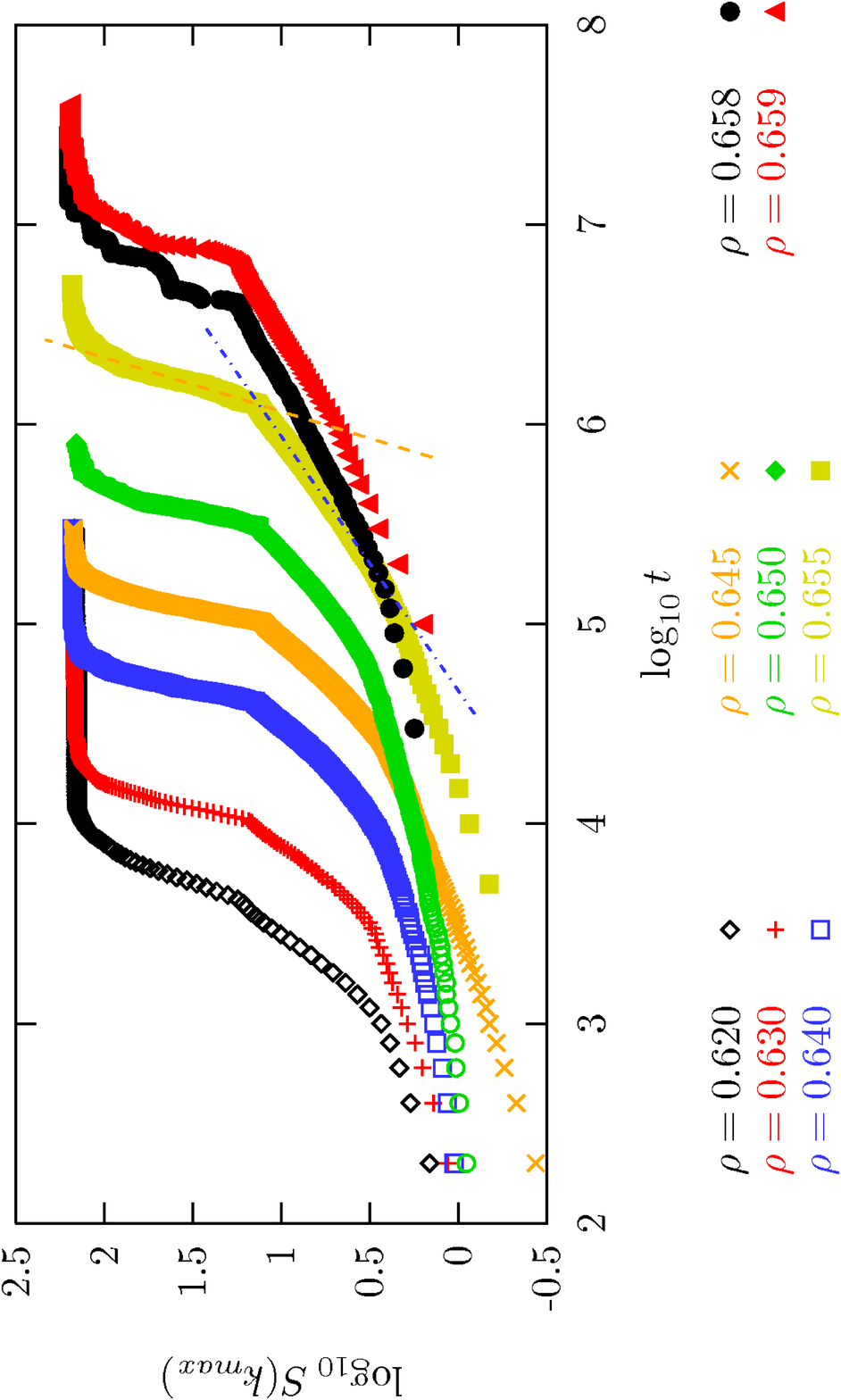}
\end{center}%
	\caption{Highest peaks of the structure factor versus time for several densities lower than the hypothetical arrest transition at $\rho=2/3$.
	Starting from a fluid state equilibrated at high $T$, we fix the temperature at $T=0.4$ (so that the equilibrium state is the crystal).
	As an example, for $\rho=0.655$ fits for two stages of crystallization have been calculated.
	Power laws $t^{\alpha}$ with exponents $\alpha=0.79$ and $\alpha=3.71$ describe first stage of nucleation and rapid growth to the crystal, respectively.
    Increasing the density, the system crystallizes less and less easily: both induction and crossover times increase.
	}
	\label{fig:sk_averaged_crystal_region}
\end{figure}
In Fig.~\ref{fig:sk_averaged_crystal_region}, the highest peak of the static structure factor $S(k_{max})$ is plotted against time for low densities, i.e. values of $\rho$ for which the crystalline state is accessible within our simulation times.
Comparing the plot with experimental results \cite{schatzel1993}, analogies and differences emerge.
In  experiments, two stages are usually recognised: an initial one, characterized by an intensity growth between $t^3$ and $t^4$, and a second process with linear or sublinear behavior.
Here (Fig.~\ref{fig:sk_averaged_crystal_region}), it seems more a three-step behavior: a regime of slowly increasing $S(k_{max})$ is followed by a rapid growth and a plateau revealing that the state is fully crystalline, as it can be verified by looking at the final states.
As shown in the two fitting examples, the first process can be roughly described by sub-linear power laws, and  the subsequent growth by a strong power law $t^{\alpha}$ with $\alpha\approx4$.
This scenario, nonetheless, is compatible with classical nucleation:
crystallinity grows up to a critical value and then jumps to the full crystal.
As the melting point is $\rho_m\approx0.637$, the two lowest densities correspond to the two phase region, but the final value of $S(k_{max})$ is roughly the same as the one of higher densities, because the fluid fraction is extremely small.

\begin{figure}[htb]
\begin{center}
	\includegraphics[height=0.9\columnwidth,angle=270]{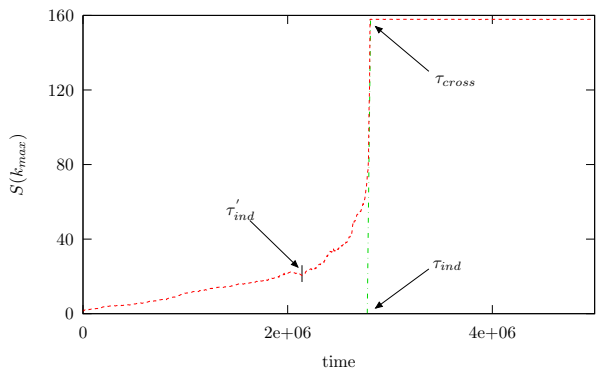}%
\end{center}
\caption{Example of a simulation for $\rho=0.655$, $T=0.4$.
The plot shows how the induction times $\tau_{ind}$ and the crossover times $\tau_{cross}$ can be determined.
$\tau'_{ind}$ represents our alternative definition (see text).
}
\label{fig:definition_char_times}
\end{figure}

As in the experiment by Harland and van Megen \cite{harland1997}, we define the \emph{induction time} $\tau_{ind}$ as the end of the nucleation process, and the \emph{crossover time} $\tau_{cross}$ as the time when the equilibrium state has been reached; the growth time is $\Delta\tau=\tau_{cross}-\tau_{ind}$ (see also Fig.~\ref{fig:definition_char_times}).
Quite interestingly, the value of $S(k_{max})$ at the induction time of Fig.~\ref{fig:sk_averaged_crystal_region} appears to be roughly the same, with a slow increase closer to arrest.
This looks slightly different from experiments, where the crystallinity seems to increase more with the volume fraction.
The explanation is probably related to the fact that the crystallinity, based on the integration of the structure factor around the interesting peak, is  a measure of the amount of order in the system and not a quantity proportional to the size of the largest crystallite.
Thus, if the density of nuclei increases with $\phi$, it is possible that the size of the critical nucleus stays roughly constant, whereas the number of nuclei increases, so that the sample exhibits a larger crystallinity, but the peak intensity remains constant.
This could also give some insight into the phenomenon showed by the experiments after a time of the order of days \cite{vanmegen1993b}, when the number of crystallites increases with the concentration, but their size gets smaller and smaller, up to dynamical arrest.
That time is well beyond the crossover time and cannot be directly inspected by our simulations, because of a finite size effect which obscures coarsening involving length scales usually much larger than the simulation samples.

Approaching the apparent arrest transition, the intensity peak at $\tau_{ind}$ starts increasing more rapidly.
The reason is that when the typical cage size becomes comparable to the critical nucleus size, nuclei can grow larger than the critical value without jumping to crystal formation and growth.
In other words, the nuclei get caged by the kinetic constraint of the model and cannot grow any more.

\begin{figure}[hbtp]
\begin{center}
\includegraphics[height=0.99\columnwidth,angle=270]{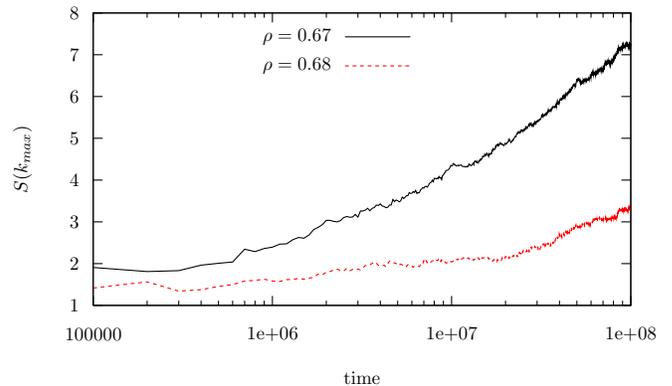}%
\end{center}%
	\caption{$S(k_{max})$ as a function of time for two densities larger than 2/3.
	Deeply into the glass region, the crystal is not observed within the simulation time.
	A very weak increase of ordered assemblies is observed, but the time scale is enormously slower than for $\rho<2/3$ (note the linear scale of the vertical axis).
	}
	\label{fig:sk_averaged_arrest_side}
\end{figure}
The situation beyond the apparent arrest transition (Fig.~\ref{fig:sk_averaged_arrest_side}) is also relatively clear, as there is no evidence of critical nucleus formation on the accessible time scale.
The question arises as to whether the arrested phase is simply a continuation of very slow nucleation beyond times accessible to simulations/experiments.
It is also possible that the absence of gravity prevents any type of sedimentation and so true arrest is never reached.

\begin{figure}[hbtp]
	\includegraphics[height=0.99\columnwidth,angle=270]{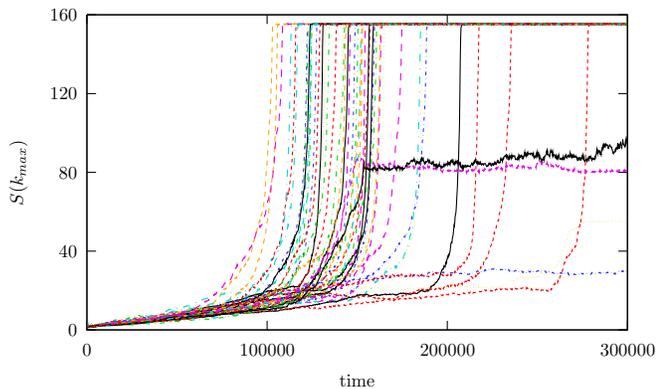}%
	\caption{
	Plots of the time evolution of the main peak of the structure factor for many simulations: 50 samples for $\rho=0.645$.
	}
	\label{fig:sk_rho_0645_many_samples}
\end{figure}

\begin{figure}[hbtp]
	\includegraphics[height=0.99\columnwidth,angle=270]{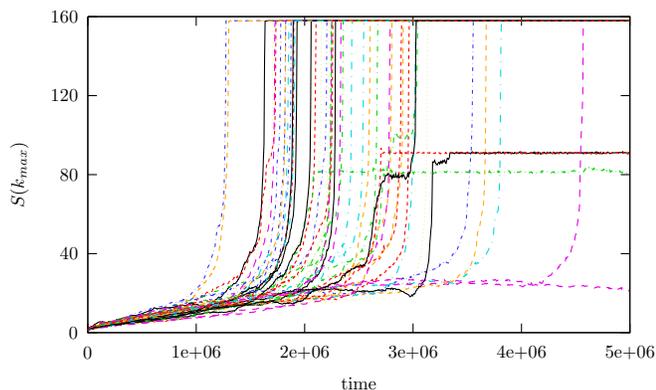}
	\caption{
	Plots of the time evolution of the main peak of the structure factor for many simulations: 50 samples for $\rho=0.655$.
	}
	\label{fig:sk_rho_0655_many_samples}
\end{figure}

In the plots of Figures \ref{fig:sk_averaged_crystal_region} and  \ref{fig:sk_averaged_arrest_side}, the structure factor was averaged over a number of different simulations.
This has been done to reproduce the global effect which should be observed on larger scales, as in the experiments.
However, this procedure actually hides the behavior  of the single simulation runs, which present an interesting behavior.
Thus, in Fig.~\ref{fig:sk_rho_0645_many_samples} and \ref{fig:sk_rho_0655_many_samples}, $S(k_{max})$ is plotted for several samples at densities $\rho=0.645$ and $\rho=0.655$.
It is quite evident that different runs have  different induction times.
This scenario is unexpected, because it is  not observed in typical first order transition models, such as the Ising model.
As one can notice by comparing the two plots, this phenomenon becomes more important for higher densities, i.e.~the spreading of the induction/crossover times increases with $\rho$.

According to the classical nucleation theory, nuclei growth is only hindered by surface tension.
As soon as a nucleus reaches the critical size, it grows quickly overcoming the unfavourable surface contribution \cite{frenkel1955}.
Here the picture is quite different.
Figures \ref{fig:sk_rho_0645_many_samples} and \ref{fig:sk_rho_0655_many_samples}, representing the time evolution of the highest peak of the structure factor at different densities, show that a new process takes place.
Such process can be easily interpreted by the language elaborated in the past years for kinetically constrained models \cite{degregorio2005,lawlor2005}.
Those models are characterized by the presence of large scale cages and a diverging dynamical correlation length, which prevents diffusion at large density.
In our model, this process interacts with nucleation and growth, because in a dense system the average cage size increases, becoming comparable to the critical nucleus size.
At this point, it becomes very probable for a crystallite to get trapped into a cage, and so it cannot grow as in the classical picture.
In Fig.~\ref{fig:sk_rho_0645_many_samples} we can single out three main behaviors.
(a) A set of runs do show a behavior which is classical: the peak of the structure factor reaches a critical value and then grows rapidly to a large crystal.
In this case, the critical nucleus is not caged and therefore it can grow without constraints.
(b) In many other cases, $S(k_{max})$ shows a quite erratic behavior. The system does crystallize, but a large amount of time is spent in a regime where there are critical nuclei which do not grow any further. The result is a considerable increase of the induction time.
This process can be interpreted as follows.
Many critical nuclei develop in cages, and shrink or disappear before the cage is broken.
Only when a critical nucleus forms in a region where it is not caged, the system can develop nucleus growth.
(c) The few cases in which the system never reaches the crystalline state are caused by the fact that there is not any connected hole allowing the correct set of movements to form a full crystal \cite{degregorio2005}.
In other words, the connected hole density is so low that a fraction of the samples does not contain any connected hole useful to move particles into the structure of the equilibrium crystal.
Finally, it is worth underlining that this caging process becomes more and more important approaching arrest.
The growth is more and more delayed and $S(k_{max})$ evolution is characterized by non-monotonic behavior (see for example Fig.~\ref{fig:example_S_star_rho0659}).

\subsection{Characteristic times}
We focus now on the distribution of induction and crossover times on approaching arrest.
Each simulation presents a quite sharp transition to the crystal.
Therefore, except for a few particular cases, the single sample induction time is almost coincident with the crossover time.
Thus, we give an alternative definition of induction time ($\tau'_{ind}$), as the instant after which the growth of $S(k_{max})$ has changed its speed and the sample clearly evolves to the crystal without large backward moves (Fig.~\ref{fig:definition_char_times}).
\begin{figure*}[!hbt]
	\begin{center}
	\begin{minipage}[b]{0.50\textwidth}
		\subfigure[$\rho=0.645$]{%
				\includegraphics[height=0.99\columnwidth,angle=270]{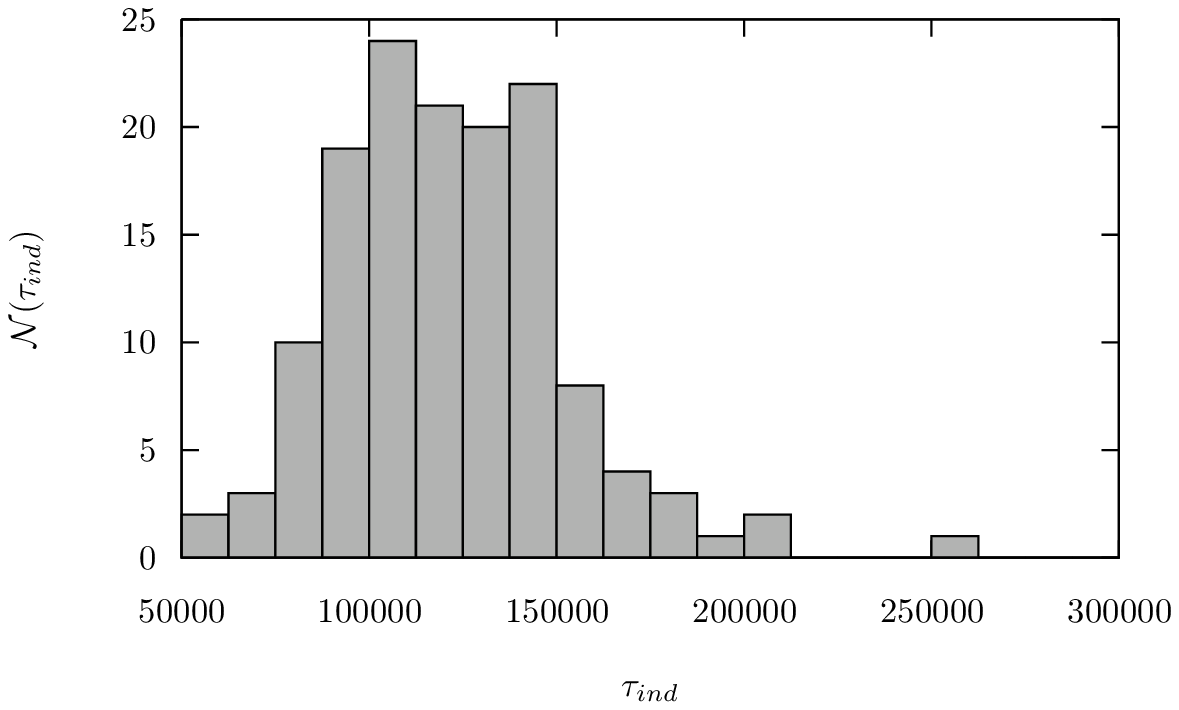}%
				\label{fig:subfig:distr_t_ind_rho0645}
		}
	\end{minipage}%
	\begin{minipage}[b]{0.50\textwidth}
		\subfigure[$\rho=0.655$]{%
				\includegraphics[height=0.99\columnwidth,angle=270]{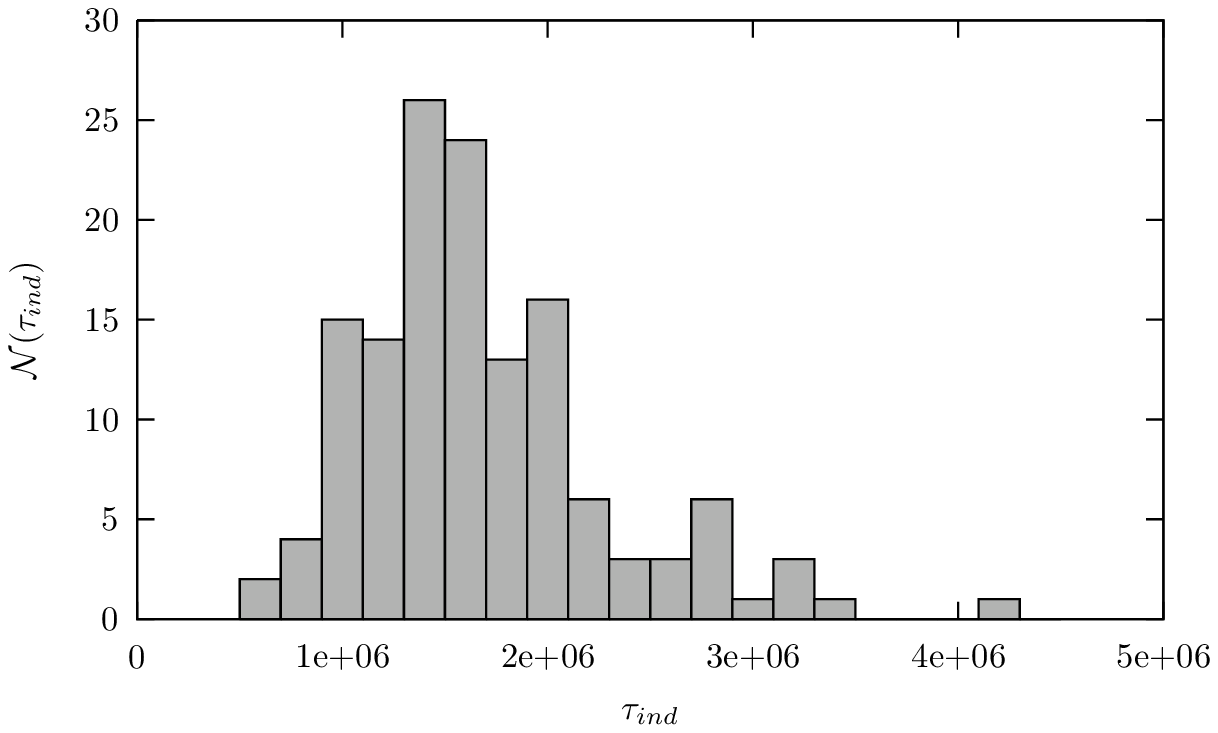}%
				\label{fig:subfig:distr_t_ind_rho0655}
		}
	\end{minipage}%
	\end{center}%
	\caption{Distribution of the induction times calculated on the time evolution of the main peak of the structure factor for 150 independent runs each. The densities chosen are $\rho=0.645$ (a) and $\rho=0.655$ (b), respectively.
	}
	\label{fig:distributions_t_ind}
\end{figure*}
\begin{figure*}[!hbt]
	\begin{center}
	\begin{minipage}[b]{0.50\textwidth}
		\subfigure[$\rho=0.645$]{%
				\includegraphics[height=0.99\columnwidth,angle=270]{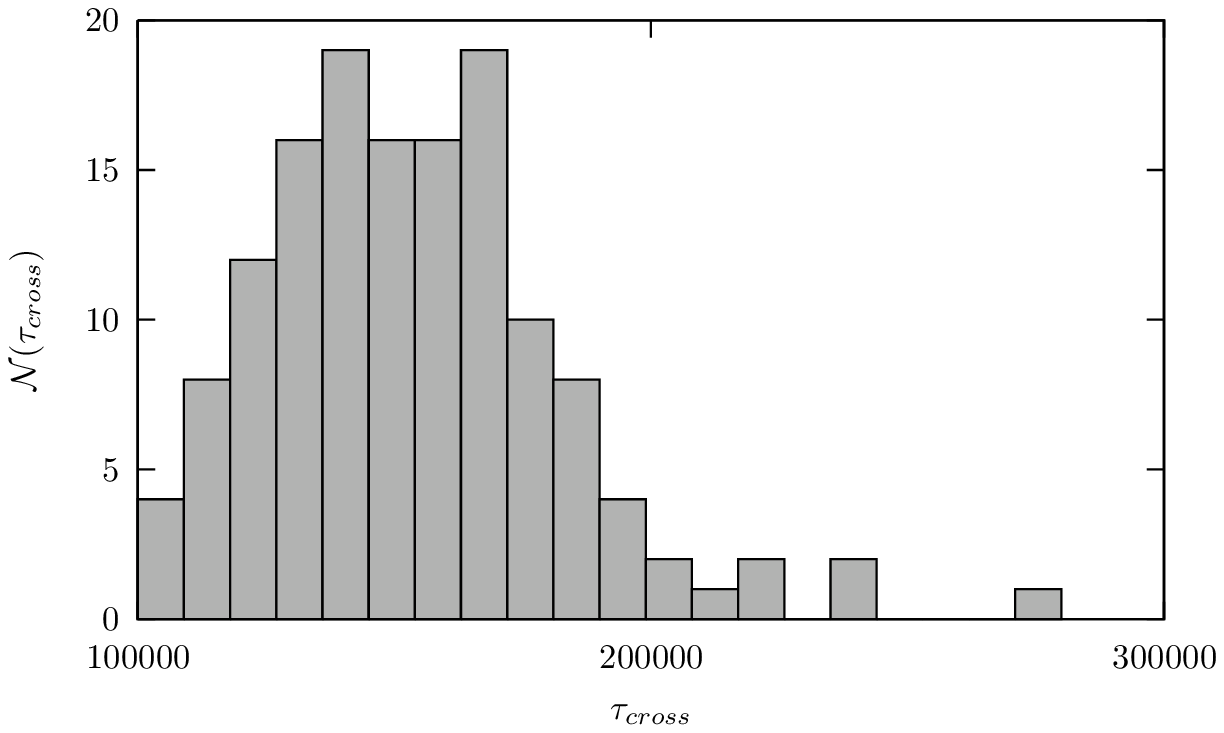}%
				\label{fig:subfig:distr_t_cross_rho0645}
		}
	\end{minipage}%
	\begin{minipage}[b]{0.50\textwidth}
		\subfigure[$\rho=0.655$]{%
				\includegraphics[height=0.99\columnwidth,angle=270]{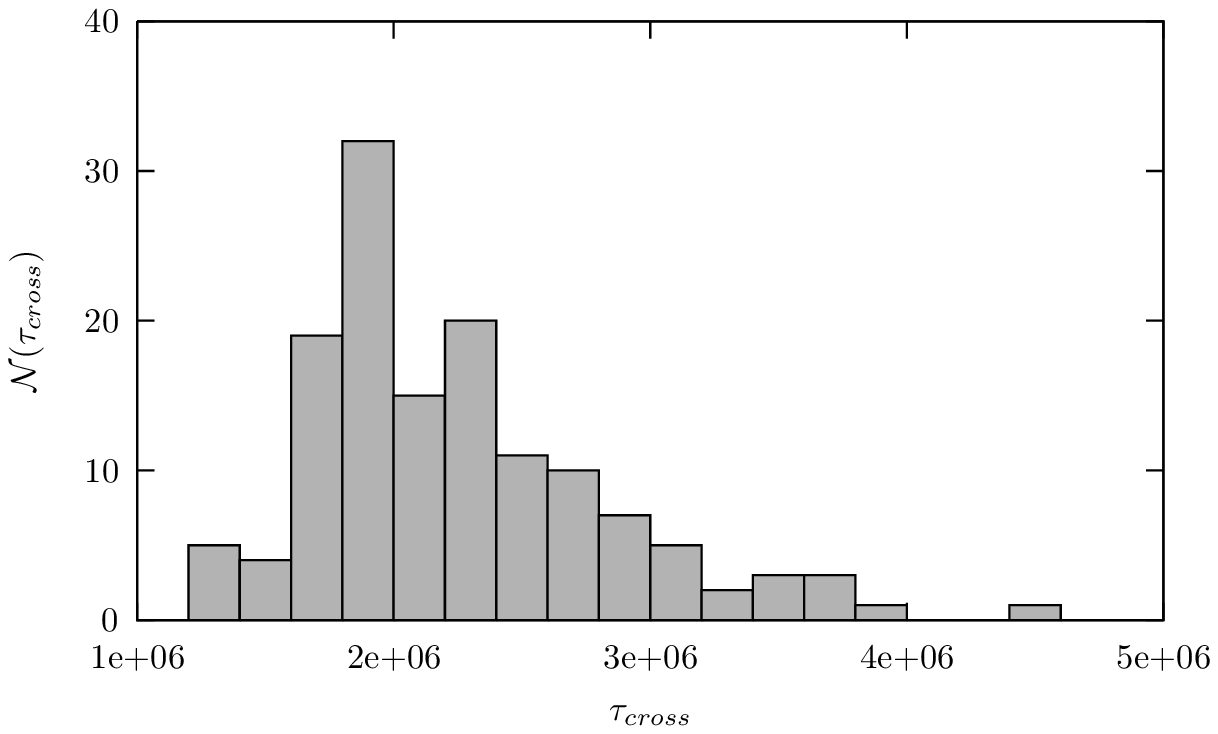}%
				\label{fig:subfig:distr_t_cross_rho0655}
		}
	\end{minipage}%
	\end{center}%
	\caption{
	Distribution of the crossover times calculated on the time evolution of the main peak of the structure factor for 150 independent runs each. The densities chosen are $\rho=0.645$ (a) and $\rho=0.655$ (b), respectively.
	}
	\label{fig:distributions_t_cross}
\end{figure*}

In Figures \ref{fig:distributions_t_ind} and \ref{fig:distributions_t_cross}, the distributions of the induction and the crossover times are illustrated  for the densities $\rho=0.645$ and $\rho=0.655$.
Given the limited number of samples (150), it is hard to make definitive statements.
However, it is possible to notice a few interesting phenomena.
For example, it is evident that the width of the distributions increases with $\rho$.
More tentatively posed, such distributions do not appear to be gaussian and seem to be skewed towards higher times.
Moreover, it is interesting to note that approaching arrest this asymmetry seems to  increase slightly towards higher times.
\begin{figure}[htbp]
\begin{center}
	\includegraphics[height=0.85\columnwidth,angle=270]{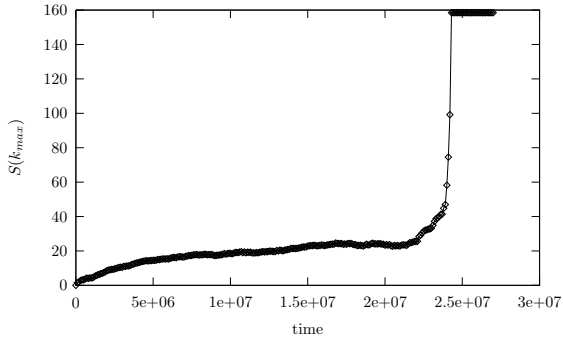}%
\end{center}
\caption{Example of the time evolution of a sample very close to dynamical arrest ($\rho=0.659$).
Peaks of $S(k)$ are plotted against time.
It is noticeable that the curve is non-monotonic.
}
\label{fig:example_S_star_rho0659}
\end{figure}

As we have seen up to here, the kinetic processes that are still within the crystal phase but very close to dynamical arrest (see for example Fig.~\ref{fig:example_S_star_rho0659}) are quite peculiar, sufficiently so that we can no longer clearly identify the conventional nucleation and growth regimes.
Instead, order grows in the system via a series of quite sudden changes, followed by periods where the system appears to be almost immobile.
Furthermore, quite unlike conventional crystallization, the peak of the structure factor is no longer monotonic, and order can at first increase, then diminish for periods of time.
This is a consequence of the wrong type of order having been formed, leading to a  sort of ``blind alley'', from which the system ultimately has to reverse, and try again to find a more successful pathway.
In contrast to the classical picture, in the vicinity of arrest many nuclei grow up to few lattice steps and then jam each other instead of growing further.
This creates a long-lived state, in which most  particles are blocked or move locally, especially in proximity to the interfaces between nuclei.
This phenomenon does not appear to be a finite size effect, because simulations at larger sizes  show similar crystallite sizes.
%
To our knowledge, this is the first time such
an intermittent mechanism has been observed in systems ordering near arrest; we suggest that these processes may be quite general, being simply related to a hard repulsive interaction and the mechanism of caging.

The induction and crossover time of colloidal particles have been reported in a few papers \cite{harland1997,heymann1997,schope2006}.
Harland and van Megen \cite{harland1997} studied the same type of particles described in the experiment by Pusey and van Megen \cite{pusey1986}, with an effective radius equals $R = 201\pm1 nm$, the polydispersity was $s\approx5$\% and the fusion and melting volume fractions were $\phi_f=0.494$, $\phi_m=0.545\pm0.003$, respectively.
Sch\"ope \textit{et al.} \cite{schope2006} studied similar particles with a radius $R = 320 nm$, polydispersity $s = 4.8$\%, and $\phi_f=0.505$, $\phi_m=0.538$; these two last values are based on the measurements of Kofke \textit{et al.} \cite{kofke1999}.
For the data by Sch\"ope \textit{et al.}, dynamical arrest is estimated to occur at $\phi=0.575$.
We now make a direct comparison between the experimental data and the results from our model.
In order to do that, it is necessary to translate the plot of our characteristic times so that the melting point of the model, $\phi_m=0.637$ coincides with the melting point for hard spheres $\phi_m=0.545$.
Then, we shrink our data on the $\rho$ axis in order to make the hypothetical arrest transition of the model $\phi_g=2/3$ coincident with the hard sphere one: $\phi_g=0.58$.
Finally, the characteristic times have been multiplied by 0.1, so that they can fit the range of the experimental data.
This is acceptable, because the simulation times do not have an absolute meaning, but they are related to real time through a multiplicative factor.
\begin{figure}[htbp]
\begin{center}
	\includegraphics[height=0.95\columnwidth,angle=270]{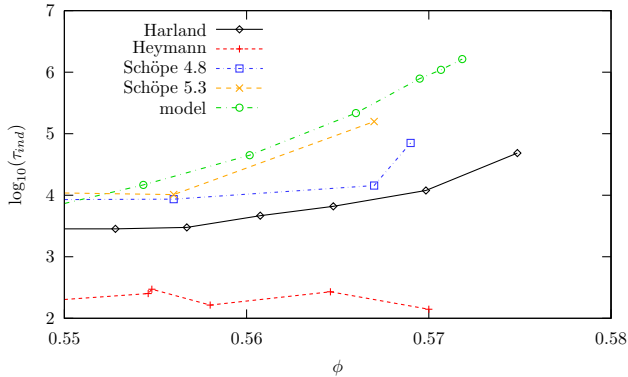}%
\end{center}
\caption{
	Comparison of plots of the induction time $\tau_{ind}$ versus density for several experiments and results from our model.
	The experimental data are taken from the papers: \cite{harland1997,heymann1997,schope2006} ($s = 4.8$\%), respectively.
	H.~J.~Sch\"ope also provided us with data for polydispersity $s = 5.3$\%.
	The values of $\tau_{ind}$ for the model have been multiplied by 0.1, so that they can fit the range of the experimental data.
	Here we focus on the asymptotic behavior close to the arrest transition, so what we compare is only the functional form of the divergence, and not the exact values of the characteristic times.
}
\label{fig:asymptotic_ind_times}
\end{figure}
\begin{figure}[htbp]
\begin{center}
	\includegraphics[height=0.95\columnwidth,angle=270]{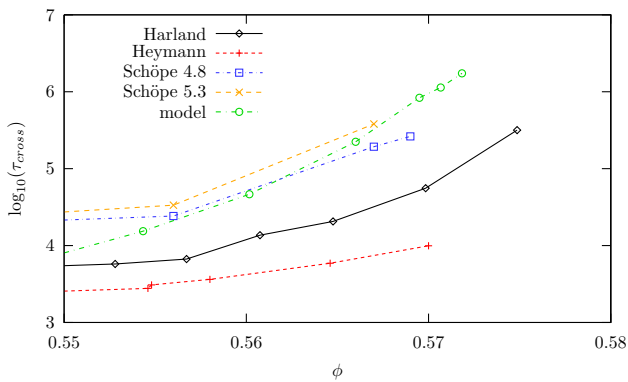}%
\end{center}
\caption{
	Comparison of plots of the crossover times versus density for several experiments and results from our model, as in Fig.~\ref{fig:asymptotic_ind_times}.
	Here we focus on the asymptotic behavior close to the arrest transition.
}
\label{fig:asymptotic_cross_times}
\end{figure}
Fig.~\ref{fig:asymptotic_ind_times} compares our rescaled induction times with sets from several experiments.
Looking at the model results, it is evident that induction times $\tau_{ind}$ grow sharply as we approach the arrest transition.
The rapid increase of the induction times shows that it becomes increasingly difficult to form a critical nucleus that would liberate enough new free volume around it to make the growth process more rapid. Even when the critical nucleus is formed, the process of rearranging the system around it becomes more difficult.
Therefore, it is apparent that the classical picture of nucleation and growth is no longer applicable.
In Fig.~\ref{fig:asymptotic_cross_times}, the comparison of the crossover times is illustrated.
Unfortunately, in both cases the picture is not very clear, and even the different experimental sets are quite heterogeneous.
The experimental procedure partially affects the data and polydispersity also has a significant effect, as can be understood by a simple comparison between the two data sets from Sch\"ope \textit{et al.}.
In spite of these weaknesses, a similar trend is recognisable at high $\phi$ in both induction and crossover times.
In particular, a comparison between the asymptotic behavior of the model and the data from Harland and van Megen (the ones closest to arrest), is intriguing.
There appears to be a quite similar functional form in the last points, a sign that could reveal an analogous divergence.
\begin{figure}[htbp]
\begin{center}
	\includegraphics[height=0.95\columnwidth,angle=270]{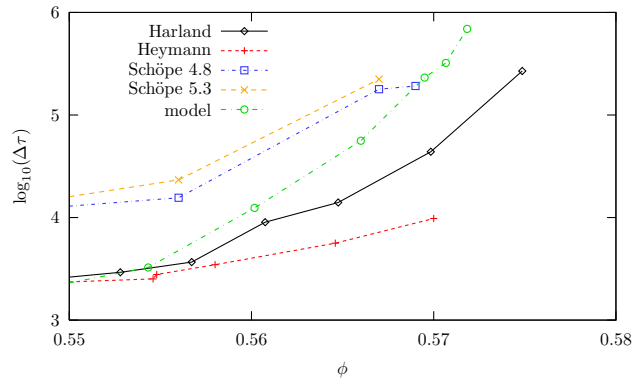}%
\end{center}
\caption{
	Growth times $\Delta\tau$ vs $\phi$ for the model and experiments as in Fig.~\ref{fig:asymptotic_ind_times}.
}
\label{fig:asymptotic_delta_t}
\end{figure}

The duration of the growth regime ($\Delta\tau$) also increases as dynamical arrest is approached.
In Fig.~\ref{fig:asymptotic_delta_t}, we present a comparison of the behavior of $\Delta\tau$ vs $\rho$ between our model and experimental data as in Figures \ref{fig:asymptotic_ind_times} and \ref{fig:asymptotic_cross_times}.
Again, our results have been processed in order to make $\phi_m$ and $\phi_g$ of the model coincident with the experimental values.
Quite remarkably, the asymptotic behavior of the model and the data from Harland and van Megen  appear to be similar on approaching arrest.
No theory for either of these laws exist as yet, though it is natural that, as overall system mobility vanishes, they would also acquire a related divergence.

\section{Conclusions}
We note that our simple lattice model has been able to reproduce a wide range of phenomena from real glasses and energy landscape models, including onset of collective behavior, divergence of the characteristic times, and many properties of crystal nucleation in the vicinity of dynamical arrest.
The reason of the good performance of the model can be interpreted as follows.
Models based on kinetic constraints alone do create a complex effective free energy
landscape in which, at sufficiently high density, many movements are prohibited by infinite or large barriers.
Nonetheless, many dynamical pathways (each mediated by a connected hole, in the language of the simple models \cite{degregorio2004,lawlor2005,degregorio2005}) involving long ranged transport still remain at fixed (zero) energy.
In such models, true dynamical arrest only occurs when the lattice is fully filled. This dramatic reduction of dynamical pathways induced by kinetic constraints certainly leads to dynamical slowing, but not to true glass behavior.
If we now allow different local energies within different cages one obtains a complex energy landscape. 
Then, rare pathways that were formerly barrier-less remain ``easy'', but acquire a multitude of smaller energy barriers. The accumulation of such bumps against the backdrop of a vanishing number of easy pathways ultimately leads to interesting singular behavior for the characteristic time, that is considered to be truly representative of the glass state.
This is the root of glass behavior, and its physical origins are quite clear in our model.
Similarly, the presence of some of these pathways allow for the formation of the crystal.

The only property that is not included in the model is the effect of polydispersity, which will be examined in a future project.
However, a recent numerical work \cite{zaccarelli2009} shows that the effect of polydispersity on the dynamics is quite small and crystallization can also occur in a monodisperse system of hard spheres.
Quite interestingly, from the analysis of our model it emerges that dynamical slowing and anomalous crystallization appear to be essentially driven by the interplay between the kinetic barrier and the underlining crystal phase, as illustrated above.

It must be considered intriguing that the ingredients of the model are already known in the literature but that they have not hitherto been combined in this way.
Purely repulsive Hamiltonian lattice models without kinetic barriers 
\cite{biroli2002,ciamarra2003a}
appear not to yield a KWW characteristic time law, as in experiments and continuous simulations.
On the other hand, purely kinetic models do not have a crystal phase
\cite{kob1993}.
Here we have a simple model that reproduces, in spite of the weaknesses implied by the discretization of the space, several main effects associated with colloidal systems. Thereby it opens up the possibility of a more transparent dialogue
between experiment, simulation and theory.

We thank P.~N.~Pusey, W.~K.~Kegel, and H.~J.~Sch\"ope for useful interactions.
This paper was written within an EU-US research
consortium, funded in part by a grant from the Marie Curie
program of the European Union (``Arrested Matter'', Contract
No.~MRTN-CT-2003-504712).

\bibliography{biblio}

\end{document}